\documentclass[10pt,a4paper]{article}

\usepackage{graphics,tikz}
\newcommand{\bea}{\begin{eqnarray}}
\newcommand{\eea}{\end{eqnarray}}
\newcommand{\be}{\begin{equation}}
\newcommand{\ee}{\end{equation}}

\def\be{\begin{eqnarray}}
\def\ee{\end{eqnarray}}
\def\bd{\begin{displaymath}}
\def\ed{\end{displaymath}}

\def\ga{\gamma}

\def\etal{{\em et al.}}
\def\ADNDT{{\em At. Data. Nucl. Data. Tables }}

\def\NP{{\em Nucl. Phys. A }}
\def\PR{{\em Phys. Rev. C }}

\def\PRL{{\em Phys. Rev. Lett. }}

\def\jpg{{\em J. Phys. G: Nucl. Part. Phys. }}
\def\EPJ{{\em Eur. Phys. J. A }}
\def\IJME{{\em Int. J. Mod. Phys. E  }}

\begin{document}


%

%
\title{Importance of Q-values in astrophysical $rp$-process}
\author{Chirashree Lahiri and G. Gangopadhyay\\
Department of Physics, University of Calcutta\\
92, Acharya Prafulla Chandra Road, Kolkata-700 009, India\\
email: ggphy@caluniv.ac.in}
\maketitle

\maketitle
\begin{abstract}
The importance of measuring Q-values in rapid proton capture process has been 
investigated. The microscopic optical model, derived using a nucleon nucleon
interaction and densities from relativistic mean field calculations,
has been utilized to calculate the reaction rates. It has been observed that
the Q-values involved in the reactions at waiting points at $A=60$ and 64 are very
important in determining the final abundance of the process.  
Some other Q-values also play a crucial role in the final
abundance of nuclei near the end point of the process.

\end{abstract}

\section{Introduction}
The astrophysical rapid proton ($rp$) process involves proton rich nuclei
at or beyond the proton drip line. Measuring nuclear masses in these nuclei 
is a very difficult problem. Even when measurements are available, they have 
very large errors in many instances. In many other nuclei, one has to depend on 
the theoretical estimates obtained from various mass formulas. In the present
work, we intend to investigate the effect of the mass uncertainty on the final 
abundance of $rp$-process. We study the possible effect of the Q-value of a particular
proton capture reaction, or equivalently, the proton separation energy $S_p$, on the final abundances at different masses.

The importance of the nuclear mass in $rp$-process lies mainly in the balance 
between the forward $(p,\gamma$) reaction and its inverse. Particularly, at the so called 
waiting points, {\em i.e.} nuclei with even $N=Z$, these two processes compete with each
other. A small positive or a negative $S_p$ usually ensures that the 
inverse process dominates. In such a scenario, the $rp$-process may get 
stalled and wait for $\beta$-decay of the waiting point nucleus. Of course, it 
is well known that at certain temperature range, depending on the Q-value of 
the reaction, two proton capture can bridge the waiting point enabling the 
$rp$-process to continue without any hindrance. The role of nuclear mass in 
bridging the waiting points has been discussed in various works [See for 
example Illiadis\cite{book}, Schatz\cite{schatz0}, or Refs.\cite{ijmpe,epja} and references 
therein]. 

The end point of the $rp$-process 
has also been investigated in 
detail\cite{schatz,ijmpe1}
under different temperature-density 
and proton fraction profiles. 

\section{Method of Calculation}

Apart from proton capture, a nucleus can also undergo decay by 
emitting beta particles
while, for higher mass isotopes, $\alpha$-decay is another probable channel. 
In this work,
the measured half life values for $\beta$-decay have been taken from the compilation
by Audi et al.\cite{audi1} except in the case of $^{65}$As, which is taken from the experimental
measurements by L\'opez et al.\cite{lopez} In absence of experimental data, half life values have
been taken from the calculation by M\"{o}ller et al.\cite{Mol} both for $\beta$- and $\alpha$-decay. This results in a set of coupled differential equations
which have been simultaneously solved to obtain the nuclear abundances as a 
function of time.  

\begin{figure}[t]
\resizebox{10cm}{!}{\includegraphics{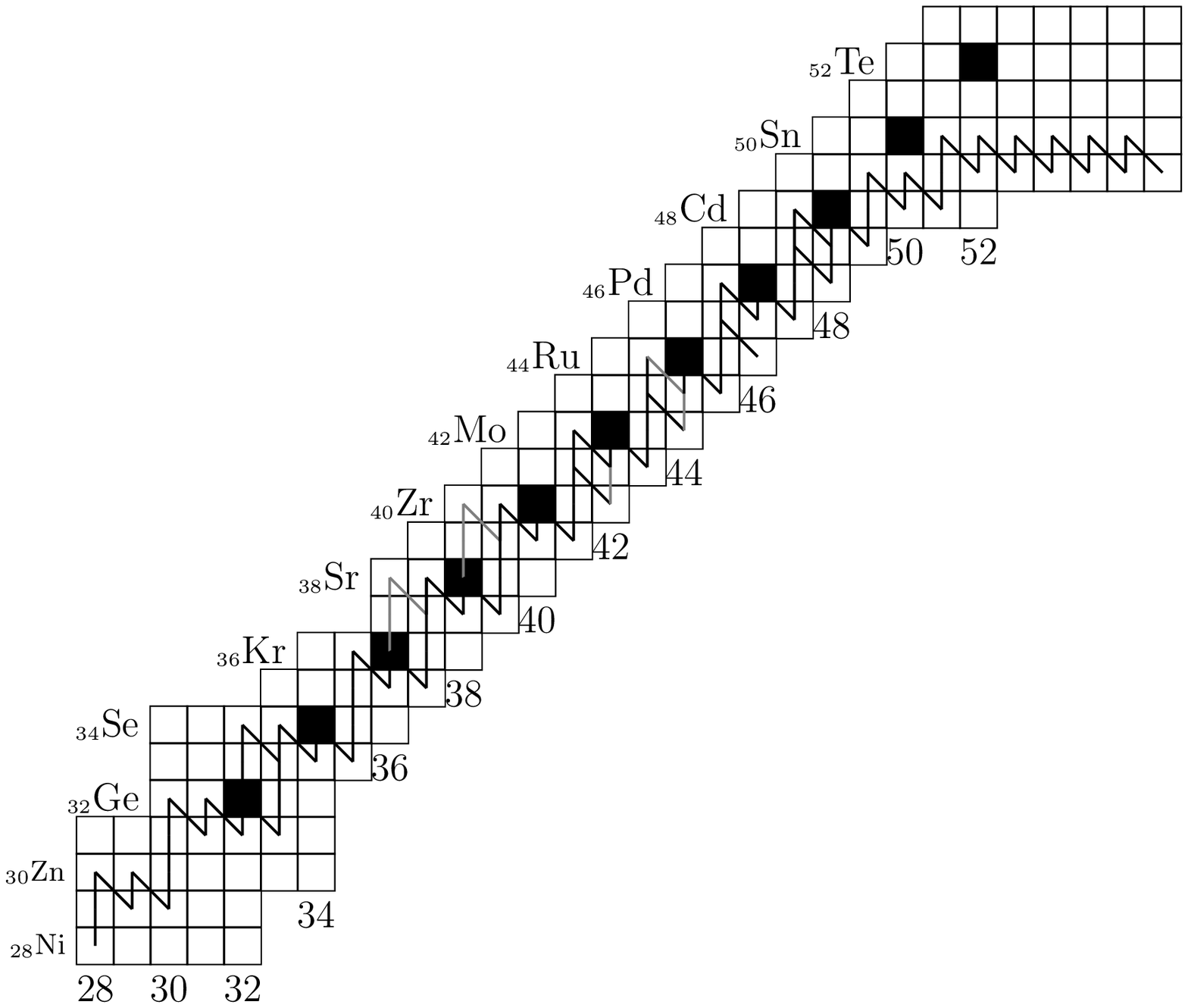}}
\vskip -6cm
\caption{\label{fig:a} The rp-process path for 1.5 GK}
\end{figure}

In the present work, we concentrate on the $(p,\gamma)$ reactions whose Q-values have either
not been determined experimentally or,  have very large errors. We look at all 
the reactions that lie on the possible path of the $rp$-process starting from 
$^{56}$Ni. There are numerous works that have calculated the above path. In Fig. 
\ref{fig:a}, we present such a path from at 1.5 GK Ref.\cite{ijmpe}. The 
dark and the light lines indicate major (flux more than 10\%) and minor (flux between 1 to 10\%) paths, 
respectively. The black boxes indicate waiting points. For this path, we assume 
that the process has a duration of 100 seconds. The density is assumed to be 
10$^6$ gram/cm$^3$ and the proton fraction is assumed to be 0.7. This 
corresponds to a scenario denoted as Model I in Lahiri \etal\cite{ijmpe1} as 
well as here. Though, this model is not very realistic, it helps us to 
understand the flow of the abundance. We will introduce a more
realistic scenario at a later stage.

The rates for the astrophysical processes have been calculated in the 
microscopic optical model using densities from relativistic mean field (RMF) 
calculations and 
density dependent M3Y (DDM3Y) potential\cite{ddm3y,ddm3y1}. Briefly, one starts with a density dependent M3Y interaction and folds it
with the nucleon density in the target to obtain a nucleon-nucleus potential. The interaction is given by

\begin{equation}
 v(r,\rho,E)=t^{M3Y}(r,E)g(\rho)                 
\end{equation}
where $E$ is incident energy and $\rho$, the nuclear density.
The $t^{M3Y}$ interaction is given by
\begin{equation}
t^{M3Y}=7999\frac{e^{-4r}}{4r}-2134\frac{e^{-2.5r}}{2.5r}+J_{00}(E)\delta(r)
\end{equation}
where $J_{00}(E)$ is the zero range pseudo potential,
\begin{equation}
J_{00}(E)=-276\left( 1-0.005\frac{E}{A}\right) {\rm MeV} fm^{3}\end{equation}
and $g(\rho)$ the density dependent  factor,
\begin{equation}
g(\rho)=C(1-b\rho^{2/3})\end{equation}

For the interaction, we have made use of the default parameters, which had been obtained from nuclear matter calculations\cite{ddm32}, without any modification.

In order to fit the experimental data, the folded DDM3Y
potential is multiplied by factors 0.7 and 0.1 to obtain the
real and imaginary parts of the optical potential, respectively. We emphasize that better fits for individual
reactions are possible by varying different parameters. But, as we are dealing with a mass region where the experimental mass
values are hardly available,
this approach is clearly inadequate. Therefore,
we have refrained from fitting individual reactions and throughout the rest of the work, we use these two factors to
obtain the potential.
These constants, along with the prescriptions
for level density and E1 strength,  have been estimated
by fitting the available S-factors for low energy proton reactions in mass 60-100 region\cite{epja,prc}.

We have calculated the Q-values (say, $Q_0$) for proton capture reactions using 
experimental masses from the compilation by Audi \etal\cite{audi} or, in their absence, the new 
phenomenological formula\cite{mass1} and looked at the variation over
the range $Q_0\pm \sigma$, where $\sigma$ has been chosen as either 
the measurement error or as equal to the rms
error of the formula, {\em i.e.} 376 keV, when experimental values are not available. 
The latter range is large enough to include almost all possible theoretical 
or systematic predictions (for example predictions from Refs.\cite{Mol,DZ}).
In the cases where the mass values have been experimentally measured with
reasonable accuracy, the effects of the uncertainties of the measurements
on the flow of the $rp$ process are insignificant. In comparison,
if the mass value is either available from theory, 
phenomenology or systematics, 
or has a large error,
the large possible variation of
mass may significantly alter the flow and the end point of the process.
We seek to identify the points on the $rp$-process path where the uncertainties of the 
Q-values of the reaction may have significant effects on the final abundance.

\section{Results}

\begin{figure}[ht]
\center
\vskip -3cm
\resizebox{12cm}{!}{\includegraphics{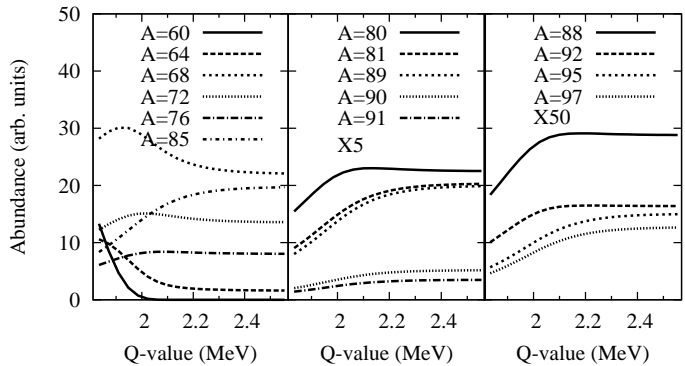}}
\caption{Abundances as a function of the Q-value of the reaction $^{61}$Ga($p,\ga$) for a constant
temperature 1.5 GK.
The abundances have been multiplied by 5 and 50
in the second and the third frames, respectively.\label{fig:2}}
\end{figure}
As an example of the variation of the final abundance, in Fig. 2, we show the 
change in abundance as a function of the Q-value of the 
reaction $^{61}$Ga($p,\ga)^{62}$Ge. The experimental value for this reaction
has not been measured.
 We plot the abundance of those  mass values 
whose final abundances  are at least 1\% of the initial seed.
In Fig. 3, we plot the corresponding data for the $^{65}$As($p,\ga)^{66}$Se
reaction. Here the experimental values, as indicated in Schury \etal \cite{ge2}
have an error of 200 keV. However, we note that in the same reference, the Q-value for proton
capture reaction in $^{64}$Ge has been measured to be -255 (104) keV.
A recent measurement\cite{ge} has measured a slightly different value, 
{\em i.e.} -90 (85) keV, which we also assume in our calculation. Another 
reference\cite{ge1} found this Q-value to be
0.401(530) keV. Though we have not assumed this value, we note that it,  
being positive, alters the flow considerably. The effect of this value,
as well as the large error, on the proton separation energy in $^{66}$Se 
also strongly affects the flow.

\begin{figure}[hb]
\center
\vskip -3cm
\resizebox{12cm}{!}{\includegraphics{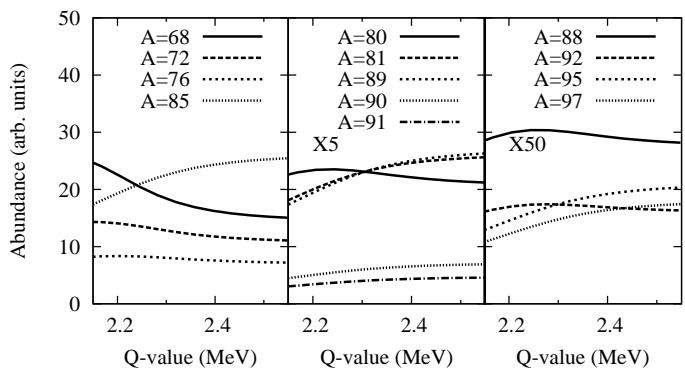}}
\caption{Abundances as a function of the Q-value of the reaction 
$^{65}$As($p,\ga$) for a constant temperature 1.5 GK.
The abundances have been multiplied by 5 and 50
in the second and the third frames, respectively.\label{fig:3}}
\end{figure}

\begin{figure}[t]
\center
\resizebox{7cm}{!}{\includegraphics{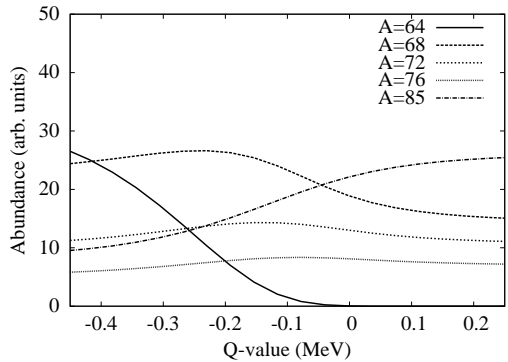}}
\caption{\label{fig:4} 
Major abundances as a function of the Q-value of the reaction 
$^{64}$Ge($p,\ga$) for a constant temperature 1.5 GK.}
\end{figure}

We further note that at the waiting points, the variation of the final product for 
changes in the Q-value of the $_Z^AX(p,\gamma)$ and the 
$_{Z+1}^{A+1}Y(p,\gamma)$ reactions can be made nearly identical by simply
shifting the origin.  For example, in Fig. \ref{fig:4} we plot the major 
abundances against variation of  the Q-value of the reaction
$^{64}$Ge($p,\ga$). It is clear that the major abundances in Fig. \ref{fig:3}
and Fig. \ref{fig:4} follow the same pattern after a shift in the origin. The 
fact can easily be explained 
as the probability 
of crossing the waiting point depends on the Q-values of both the 
$^{64}$Ge($p,\ga$) and $^{65}$As($p,\ga$) reactions. Thus,
an increase in either of them inhibits the photodisintegration in a similar way.
Hence, we have studied the dependence on only one of the relevant Q-values. 

In both the reactions, we have assumed a profile of the X-ray burster 
as given in Model I.  At all other waiting points,
the reaction Q-values have smaller effects on the final abundance. 

A few important points can be derived from the Figs. 2 and 3.
One can see that in $A=68$, and to some smaller extent $A=72$ and 76, 
{\em i.e.} the waiting points of the process, the abundance tends to fall away 
about some median value. However, $A=64$ does not behave 
at all like a waiting point. As indicated in Fig. 4 of Ref.\cite{epja}, this 
is due to the fact that at 1.5 GK,  the two proton capture process has a very
large cross-section for this nucleus.
In $A=88$ and 92, we observe a somewhat similar behaviour, though the falling off at higher $S_p$ values is very slow.
 
The population at the waiting points are not affected to a large extent
by processes that do not directly include the nuclei. Thus, for example,
the population at $A=76$ changes typically by factor of less than 1.5 over the range
of Q-values chosen for the capture reactions at $^{61}$Ga and $^{65}$As but
changes to a greater extent when the proton separation energies of 
$^{77}$Y or
$^{78}$Zr are varied. Again the waiting point at $A=64$ is an exception.
Beyond $A=76$, the 
$S_p$ values at the waiting points do not have any significant effect on the
final abundance. 
As is evident from Fig. 1, the waiting points beyond $A=68$,
except to a small extend at $A= 72$ and 76, are not bridged effectively by the $rp$-process.
Hence, these $S_p$ values are almost irrelevant to the final abundance.
For proton capture by $^{68}$Se, and to a slightly less extent in $^{76}$Sr, 
the variation in mass affects the final abundance to a very 
small extent. This 
points to the fact that these waiting points are not bridged to any 
significant extent by proton capture reactions at the conditions assumed
by us  and $\beta$-decay dominates irrespective of the $S_p$ value. 
On the other hand, Q-values for targets such as 
$^{86}$Tc, $^{89}$Ru, $^{90}$Rh and $^{93}$Pd affect the process,
indicating that the $rp$-process path shifts to these more stable nuclei
beyond $A=76$. 

We also find that the change in abundance at the waiting points, for changes 
in the Q-value of a particular reaction, tend to be closer to each other.
At masses that do not contain a waiting point, the variation is larger
and again seem to form a cluster of values near to each other.
This is due to the enhancement of a particular pathway. All the nuclei
formed in that pathway tend to vary in a similar fashion.
For example, the ratio of the final abundances of $A=91, 93, 95$ and 97,
when plotted against the Q-value of the reaction $^{86}$Tc($p,\gamma$),
remains almost a constant. This indicates that that there is effectively
 a single path leading from $^{86}$Tc to these masses. 

To explain the variation in abundance as seen in Figs. 2 and 3 further, we note 
that the flux through the waiting points are usually significantly larger than at nearby masses. If a change in the
$S_p$ value results in transfer of even a small amount of this flux to the 
beta decay channel, the flux at corresponding nuclei in non-waiting point 
masses tends to increase by a significant fraction. Once the flux at such 
a nucleus increases, it reinforces
the subsidiary paths. Hence, the nuclei that are produced, not through 
the main path but through the subsidiary paths, tend to show a similar pattern
as a function of a particular $S_p$ value. If there is any feeding
to the $\beta$-decay channel from a waiting point further up the mass scale, 
the nuclei in a 
subsidiary path that  are fed by that channel may have still
higher abundances. This feature can be seen in the case of proton capture
reactions in $^{61}$Ga and $^{65}$As. When the $S_p$ value at a particular
waiting point is very small, the 
$rp$-process gets stalled and the abundance gets trapped at the waiting point.
If the $S_p$ value is larger, the waiting point is bridged and 
flow reaches higher masses. In cases where the waiting point is bridged very
effectively, it is possible, like in mass 60 or 64, for the abundance at
the waiting point to fall to insignificant values. The abundance at the next 
waiting points then starts to fall off as the flow to these nuclei stops. 
The abundances of nuclei produced by the secondary paths, in such circumstances,
starts to level off, {\em i.e.} their rate of increase as a function of the 
Q-value of the reaction under study decreases.

If the temperature is taken to be lower, then we have a different scenario. 
The the abundance gets trapped at a lower mass value, typically $A=68$.
This ensures that the abundances at all the subsequent waiting points becomes 
constant. The $rp$-process cannot proceed very far. In fact, beyond $A=80$, no 
mass reaches an abundance of more than 1\% of the initial flux.

Finally, in a more realistic situation, where the temperature, density and 
proton fraction changes with time, we have investigated the effect of the 
uncertainties in different $S_p$ values. We have selected the time variation of 
the above quantities following Illiadis\cite{book}.
In it,  nuclear burning starts with temperature and density values of $T$ = 0.4 
GK and $\rho = 10^6$ gm/cm$^3$. After 4 seconds, the system reaches a maximum
temperature of 1.36 GK and a minimum density of $5 \times 10^5$ gm/cm$^3$. 
After 100 seconds, the temperature drops to 0.7 GK and the
density increases to $1.4 \times 10^6$ gm/cm$^3$. 
The proton fraction decreases to 0.16 at 100 seconds. This corresponds to 
Model IV of Ref.\cite{ijmpe1} where one can also see the final abundances
for the $S_p$ values predicted by Ref.\cite{mass1} for various temperature density 
profiles. 

In this particular work, we are interested in the change in 
final abundances due to variation in $S_p$ values. In Fig. 5, we study the change 
in abundances for variation of the Q-value in $^{65}$As($p,\ga$) reaction. The solid (dashed)
line provides the ratio of the final abundances between the values for 
$Q_0+\sigma (-\sigma)$ and $Q_0$. 
We find that
the baselines vary widely, particularly after 
$A=88$. This can be explained from the fact that, in a realistic model, as the 
nucleosynthesis progresses, both the temperature and the proton fraction 
decreases. Hence, the rate of rp-process nucleosynthesis slows down
towards the end. In case the $S_p$ value at $^{65}$As becomes large, 
the two proton capture cross section becomes large and the waiting point 
at $A=64$ does not hinder the process appreciably. Hence it is possible for 
nucleosynthesis to  reach $A\sim 90$ much earlier and the abundances in
$A\ge 90$ region increase substantially while those in the lower mass region
decreases. The opposite scenario occurs 
when the $S_p$ value at $^{65}$As becomes small. That is the reason for
he large variation of the baseline.
In contrast,  Model I shows sharp changes at certain mass 
numbers with a slow variation of the baseline. However, this appears unphysical
as the behaviuour is  due to 
the fact that at 1.5 GK, the change in the effective lifetime of $^{64}$Ge
due to change in $S_p$ value is larger compared to the values at lower temperatures.
This can be observed from Figure 5 of Lahiri \etal\cite{epja}. 
In Model IV the temperature,  when 
$rp$-process reaches the waiting point at $^{64}$Ge, is much lower and the
change in lifetime due to change in $S_p$ value is much less.

\begin{figure}
\center
\resizebox{8cm}{!}{\includegraphics{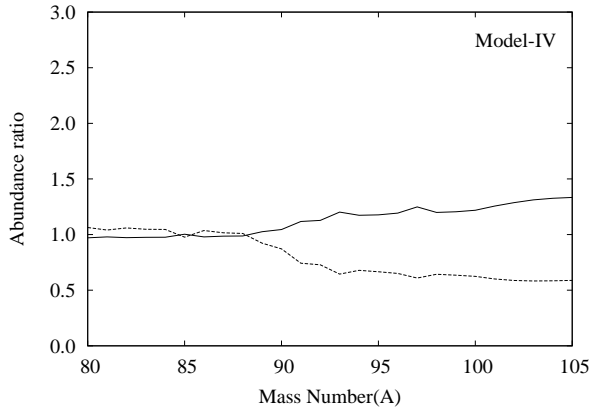}}
\caption{Ratio of change in final abundance for $Q_0\pm\sigma$ values of the 
$^{65}$As($p,\gamma$) reaction. See text for details.}
\end{figure}

To look at the existing experimental mass measurement, we find that the 
following masses, which have either not been measured  at all or measured 
with insufficient accuracy, are likely to play an important role in determining the final abundances. At the waiting points, masses of $^{62}$Ge, $^{65}$As and 
$^{66}$Se are important. Experimental values for the last two are 
available\cite{ge}, but as already noted, there is some uncertainty since 
another measurement\cite{ge1} suggests a different value for the $S_p$ value 
in $^{65}$As. Fig. 4 points out that the end point will be quiet different
in case the latter value is adopted. 
 Besides these, masses of $^{86}$Tc, $^{87,89}$Ru,
 $^{90}$Rh, $^{91,93}$Pd and $^{94}$Ag are important for
the pathways involving more stable nuclei.

\section{Summary}

To summarize, the uncertainties in the $rp$-process abundance due to the errors
in Q-values in proton capture reactions have been studied in the  
microscopic optical model. We have used the DDM3Y nucleon nucleon
interaction and densities from RMF to construct the optical potential. 
Of the waiting points, we have found that the Q-values involved in the 
reactions at masses $A=60$ and 64 are very
important in determining the final abundance of the process.  
Measurements of Q-values of proton capture reactions involving
$^{86}$Tc, $^{89}$Ru,$^{90}$Rh and $^{93}$Pd are also important in determining
the abundance near the end point of the process. 

\section*{Acknowledgment}

This work has been carried out with financial assistance of the UGC sponsored
DRS Programme and the computer facilities of the DST-FIST program of the Department of Physics of the University of Calcutta.
GG acknowledges the facilities provided 
under the ICTP Associateship Programme
by ECT*, Trento, where a part of the work was carried out.
CL acknowledges the grant of a fellowship awarded by the UGC.

\end{document}